# An alternative use of the NetLogo modeling environment, where the student "thinks" and "acts" like an Agent, in order to teach concepts of Ecology.

Aristotelis Gkiolmas[1], Anthimos Chalkidis[1], Maria Papaconstantinou[2], Zafar Iqbal[3] & Constantine Skordoulis[1]
agkiolm@primedu.uoa.gr, achalkid@gmail.com, mpapakonstan@hotmail.com, ziqbal@dangerousbend.com, kostas4skordoulis@gmail.com .
[1] Department of Primary Education, University of Athens, Greece.
[2] Department of Informatics, Ionian University, Corfu, Greece.
[3] Software Engineer, Athens, Greece.

**Abstract**

The Multi-Agent-Based programming, modeling and simulation environment of NetLogo (Wilensky, 1999) has been used extensively during the last fifteen years for educational – among other – purposes. The learning subject, upon interacting with the User's Interface of NetLogo, can easily study properties of the simulated natural systems, as well as observe the latter's response, when altering their parameters. In this research, NetLogo was used under the perspective that the learning subject (student or prospective teacher) interacts with the model in a deeper way, obtaining the "role" of an "agent". This is ***not*** achieved by obliging the learner to program (write NetLogo code) but by interviewing them, together with "applying" the choices that he/she makes on the model. The scheme was carried out, as part of a broader research, with interviews, and web-page-like interface menu selections, in a sample of 17 University students in Athens (prospective Primary School teachers) and the results were judged as encouraging. At a further stage, the computers were set as a network, where all the agents performed together. In this way the learners could watch onscreen the overall outcome of their choices and actions on the modeled ecosystem. This seems to open a new – small – area of research in NetLogo educational applications.

## Introduction

It has been noticed in the scientific literature concerning NetLogo (Wilensky, 1999), that it is an environment appropriate for teaching Complex Systems in nature (Tisue & Wilensky, 2004). In particular, it has been used to teach Complexity aspects of ecological systems and ecosystems (Vattam et al, 2011; Basu et al., 2011). All the educational researchers using NetLogo to teach about natural systems, have been using it to teach, either as a simulation environment combined with worksheets (Levy & Wilensky, 2008; 2011; Thompson, 2007) or they have previously taught their learners basic things about NetLogo programming, and then used it as an instructional tool (Hashem & Mioduser, 2011; 2012). What was considered as a relatively new approach here is that, one the one hand the student "builds" a new artifact – a NetLogo model, and simultaneously he/she learns through it, in accordance with the basic principles of Constructionism (Papert, 1991; 1993) but with no necessity of NetLogo programming when building the models. This is done by menu selections (buttons) mainly, or by oral answers. Of course, the interaction with the models (sliders, switches and buttons moved) always remains a key-feature of learning, since these are simple NetLogo models, existing already in previous Logo-like environments like StarLogo (Resnick, 1994).

The scope of the research, thus, is twofold: a) to investigate the extent in which the learners learn basic properties of ecosystems, by slightly affecting and/or interacting with simple models of NetLogo and b) to help the learners "act" like agents, building therefore partly an ecosystem of their own. The method of the research is: interviewing the students in pairs, with simultaneous interaction with the models on the computer. At the same time, they make choices – at certain stages – from menus, about "how the modeled system performs" or "how they think it should perform".

The overall teaching and learning environment here resembles Wilensky's participatory simulations with "HubNet" (Wilensky & Stroup 1999; 2000), but does not require such heavy technological equipment, since the latter does not exist in most Greek computer classrooms.

## Sample and Method

The sample of this research consisted of 17 students of the department of Primary Education, University of Athens, Greece. The students had been following the optional (but pre-requisite for some fields) topic: "Science and Environment: A Laboratory Approach", during the fall semester of 2011-2012. These students belonged to a broader sample of 85 students, which was used for research purposes, in order to study the extent to which NetLogo can help learners to become acquainted and to conceptualise systems in Nature – and mainly Ecosystems – as Complex Systems, as well as to learn basic properties of such systems.

The 17 students were sitting in front of a PC in pairs – or the last three in a triad – and were interviewed from the first of the authors for about one hour and a half. The interviews were semi – structured interviews, and were conducted both with the use of worksheets and with the use of the PC, where the three mentioned NetLogo Models were installed. Each student was giving oral answers recorded by the researcher in a digital recorder and was also completing some answers on the worksheet. There were also intervening PowerPoint slides – with menus - on the screen, where the students made – each time - specific "choices" from "menu" buttons and decided what the "behavior" of the agent(s) should be in the next execution of the model.

The NetLogo models used in order for the students to interact with them, are variations of two basic models of the NetLogo Models' Library: a) "Fire" (Wilensky, 1997a) and b) "Ants" (Wilensky, 1997b). In Figure 1a, a screenshot of the Model "Fire" is shown, and in Figure 1b, a screenshot of the Model "Ants" is depicted.

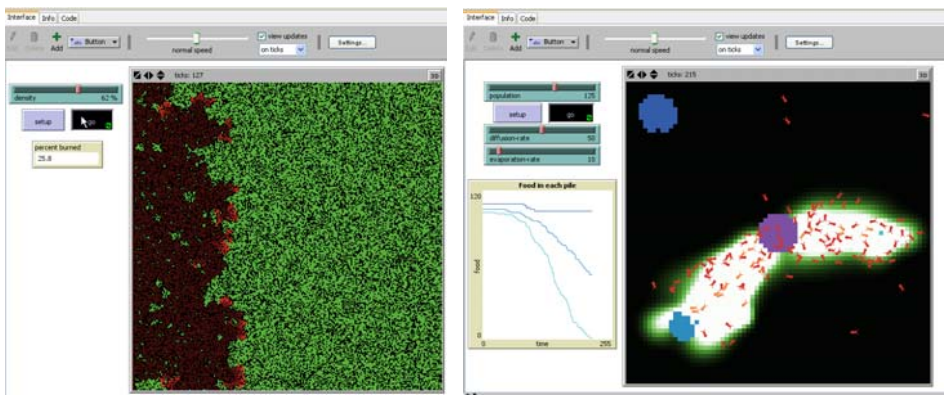

**Figure 1: a. A screenshot of the NetLogo Model "Fire". b. A screenshot of the NetLogo Model "Ants".**

The core scheme of this part of the semi-structured interviews consisted of nine questions. The first four of them (Q-F1 to Q-F4) referred to the model "Fire" and its alterations created by the authors:

Q-F1: "What do you think the exact algorithm followed by one "flame" as it moves through the patches is"?

Q-F2: "Now, what do you think the algorithm followed by one "flame" is?" (*Two variations* of the model)

Q-F3: "What would you propose as an algorithm for the "flame", in its motion on the micro-world, to have a more realistic model"?

Q-F4: "In order to have an overall more realistic representation of a forest suffering from a fire, what changes would you suggest in the model's structure?" [Afterwards, some of these changes are embedded in the model, through button selections, that the students make on a browser-like page]

The following five – main - questions in the semi-structured interview (Q-A1 to QA-5) referred to the NetLogo model "Ants", as well as variations of the model, created by the authors:

Q-A1: "What rules do you think that govern the motion of one ant when it comes out of the nest?"

Q-A2: "What would *you* choose as an ant's algorithm of moving, when it comes out of the nest, in order to have a more natural-like model?"

Q-A3: "What is the factor that guides the ant back to the nest, after it has collected one piece of food?

A. Try to answer by running the model with *only one* ant.

B. Try to answer by running the model with *very few* ants (4 or 5)

Q-A4: "Through which algorithm do the ants form the pheromone chemical trails (white trails?), in your opinion?"

Q-A5: "As far as questions Q-A3 *and* Q-A4 are concerned, what would you suggest as algorithms: (i) for one ant returning to nest through food collection and (ii) for the ants' communication through pheromone, in order to have a more realistic model? In some cases of your suggestions, you can see the model *with them included,* running on the screen.

## Results and Discussion

All the answers of the students were grouped, as regards their conceptual content, in certain categories.

In **Q-F1**:
- 12 out of the 17 students: failed totally to find the algorithm followed by the agent-"flame" as it moves through the patches.
- 3 students: argued that the flame would burn all the neighbouring patches that are "green".
- 2 students: approached closely the "correct" NetLogo algorithm for the flame, in saying that the flame burns the green patches (if any), ahead, above and below it.

But when the variation of the model was presented to them, as it is depicted in Figure 2a and then, they passed to the next question Q-F2

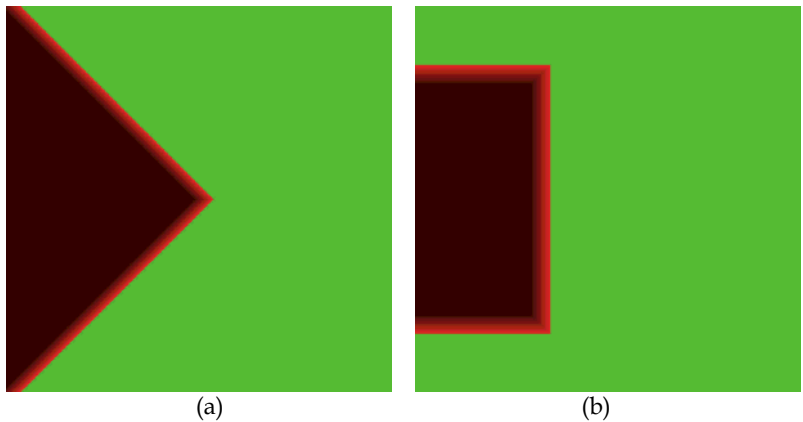

(a)                                          (b)

**Figure 2. a: A simplified variation of "Fire", where there is only one flame, starting from the middle of the left end of the Micro-World, and "burning" only up, down, left and right neighbouring patches.
b: A simplified variation of "Fire", where there is only one flame, starting from the middle of the left end of the Micro-World, and "burning" up, down, left and right but also diagonal neighbouring patches.**

In **Q-F2**
- 11 out of the 17 students of the sample: realised what the exact algorithm followed by the agent-"flame" in this model is,
- 6 of the students: still gave incorrect descriptions.

In the variation of the "Fire" Model of which a description is given in Figure 2b, the students answered that the flame "burns" all neighbouring green patches (the correct answer), to a proportion: **14/17**, whereas they gave other, not "correct" answers, to a proportion: **3/17**

**Q-F3**

Afterwards the students were entering a "web-page-like" interface where they chose, among a buttons' menu what "algorithm" they would propose for the flame as it moves from each patch to the neighbouring ones. Their given choices were:
- a. Move forward one patch regardless if the patch in front is "green" (with tree), "black" (without tree) or "red" (burnt)
- b. Move to the ones of your *three* patches ahead (front, upper front diagonal, lower front diagonal) which are "green"
- c. Move to the ones of your *five* neighbouring patches (front, upper front diagonal, lower front diagonal, above, below) which are green.

According to which one of the buttons the students were pressing, a corresponding variation of the model "Fire" of NetLogo run in front of them. After they observed and interacted with each of the three simulations, they gave oral answers about which of the three variations of the NetLogo model that they chose was closer to a real-like situation in an ecosystem. Again the results are, related, of course, to question Q-F3:
- 3 out of the 17 interviewees said that: the choice "a" created a more realistic model
- 4 out of the 17 interviewees implied that: the choice "b" created a more realistic model
- 10 out of the 17 interviewees implied that: the choice "c" created a more realistic model.

Analysing the results up to this point, there is evidence, that simplified – aphaeretic – versions of the "Fire" Model would possibly help the student "think" like an agent, in the sense of analyzing what exactly is the algorithm behind the motion of one agent. This is an apparent reason of the increase of the proportion of "correct" answers between Q-F1 and Q-F2. As a result, this understanding of the rules governing the motion of the agent in the Micro-World, lead to improved "model-planning" skills, since the answers in Q-F3 converged significantly to the correct answer ("c").

**Q-F4**

Now when it comes to Q-F4, the students were prompted to improve their "acting-like-agent" and "model-creating" skills, when they were asked to suggest improvements in the basic model, to make it more real-like. In a – similar to Q-F3 – web-page-like format, they could actually "see" and interact with the model created by their choices (they pressed buttons from a menu and an according variation of the model run). It is encouraging that many answers of the students in the interview question Q-F4, suggested correct - in their conceptual content - improvements of the model:

- 6 out of 17 interviewees suggested that: the starting point of the fire should be in the center of the Micro-world, not its left end,
- 9 out of 17 interviewees suggested that: wind direction should play a role in the model and be introduced by the user,
- 7 out of 17 interviewees suggested that: humidity (high, low or medium) should also be an important factor.

[The answers of course summed up to more than 17 because some faculty students introduced *more than one* correct improvements to the model].

In Figure 3, an overall version of the Fire model, with all the basic improvements – added to it by the students – is shown. In further research, other improvements will – hopefully – be added.

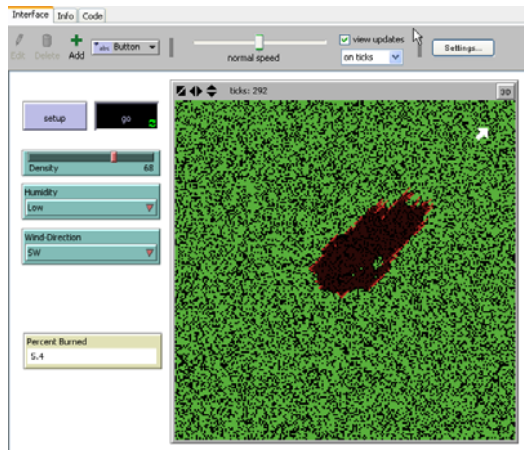

**Figure 3: A variation of the Fire Model, with spark in the center, wind and humidity.**

Regarding the NetLogo model "Ants" and its variation, related to the interview questions Q-A1 to Q-A5:

In **Q-A1**:

- 12 out of the 17 students, by content analysis of their answers: seemed to understand the random motion of the ant at each of its steps
- 3 students: tended to believe that the ant moves in a direct sense (radially), as it exits the nest from a random position,
- 2 students: gave irrelevant answers.

It is concluded by this set of results, that the students have already been able, at a certain extent, to *conceptualise the behaviour* of the NetLogo agents.

Now, passing to **Q-A2**, the students were prompted to "think" and "act" as agents themselves and suggest manners for the ants' motion. The two prevailing answers were:
- 9 students suggested: that the ants move radially but, turn to pheromone if it appears in a neighbouring patch, and
- 4 students suggested: that the ants move randomly until they find another ant that carries food and then they move radially on the direction that this ant comes from.

Again, through two buttons in a button-menu on a separate screen, the students could call the execution of the corresponding variation of the NetLogo "Ants" model, and they see that their choices indeed lead to a model very close to reality, especially the first choice.

Now as depicted in Figure 4, the students run a simplified version of the model, either with one or with few ants, and with one food pile to answer Q-A3.

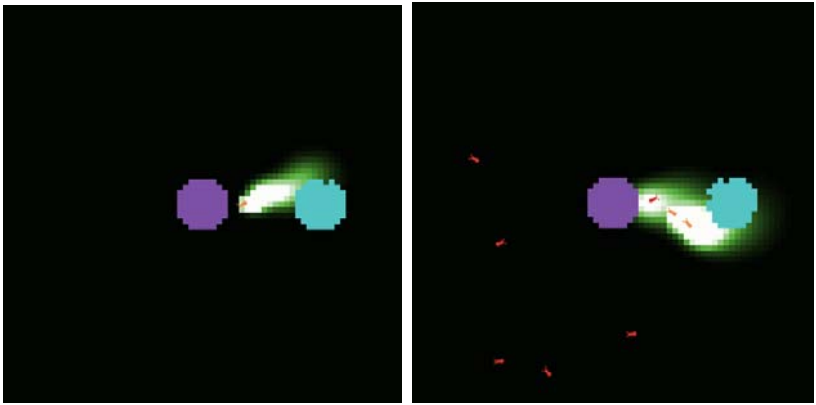

**Figure 4: Screenshots of two simplified variations of "Ants", (i) one with only one ant and (ii) one with a few ants, where the ants return to the nest after having collected one or more pieces of food. For simplicity, there is only one food pile.**

In **Q-A3**, the results were *not so encouraging*, in the sense that:
- Only 4 out of the 17 interviewees: could describe that there is a built-in algorithm in the basic model, through which the nest emits pheromone, thus leading the ants back to it, when they have collected a piece of food,
- 8 of the interviewees: stated that after the ant has collected a piece of food makes a 180 degree turn and returns to    the nest (which is wrong)
- 4 of the interviewees: could not answer or gave irrelevant answers.

In **Q-A4** which is again a question of understanding and conceptualising the agent behavior, the results were again fair enough but not very good.
- 7 of the students: gave the "correct" description, that each ant turns to its nearby patch, when the latter has pheromone (it is white),

- 6 of the students: gave answers in generalities, about ants "grouping in the areas that are white, and
- 4 of the students: could not understand at all the procedure.

Naturally, question **Q-A5** refers to "acting like an agent" and the students seem to have made appropriate choices to improving the ants' motion and behavior, so as to have a more natural-like model. The main results here are – again the overall numbers do not sum up to 17, since some students made more than one suggestions/choices:

- 5 students implied that: the ants should start exiting from the nest *only after* the first ant *with* a piece of food has returned and follow its direction.
- 10 students said that: when an ant returns to the nest *with* food, it should re-exit in a 180 degrees direction with respect to the one with which it entered the nest, and:
- 6 students thought that: the pheromone in each patch should add up, and every ant should choose from its neighbouring patches, the one that has the most of pheromone, then turning to it.

Codifying the three above answers as buttons: "a", "b" and "c" in a menu within an interface similar to the one that has already been described, the students could choose one of them each time and a) see the corresponding version of the NetLogo model on screen, as well as b) interact with its parameters.

## Conclusions

The current research project – which is a spin-off of a previous broader research that lead to a PhD, about the educational use of NetLogo for teaching systems in Nature and ecosystems as Complex Systems – has a twofold purpose: a. to make the students conceptualise natural and environmental systems' models (especially Multi-Agent-based models) from "the inside", by leading them to understand the behavior of agents, without introducing them to computer programming, and b. to make the learners capable of building models of natural systems and ecosystems, not "from the scratch", but by deciding the behavior of agents, a skill that is crucial for understanding how the modeled systems will behave under different rules. By the activities described above and by the results obtained, it seems reasonable to assume that these two aims are a bit fulfilled. Compared to other attempts to teach basic properties and relations of Ecosystems with NetLogo (Vattam, et. al., 2011; Basu et. al., 2011; Hmelo-Silver, et. al., 2011) the research here achieves results based on interaction with the modeled system itself and on making choices about it.

It is also concluded, from this little case study with the use of interviews, that students might possibly, through the use of simple NetLogo models and their variations, as well as through their navigation in specifically created interfaces, learn how to *act and "think" like members* of a natural system or ecosystem and thus understand its functions and behavior. In a further stage of application, similar to Wilensky's "HubNet" (Wilensky & Stroup, 1999), it is aimed that each student can move one agent, for example through the use of a joystick, and they altogether watch the results of their combined actions on a NetLogo screen, provided that their computers are networked.

This, we believe, is a good new tactics for teaching learners and future educators, basic aspects of environmental systems and ecosystems, within computer modeling environments.